\documentstyle[emlines,12pt]{article}
\begin{document}
\textwidth 24cm
\textheight 22cm
\oddsidemargin 1.4cm
\nopagebreak[0]
\bigskip
\medskip
\begin{center}
\bf{Operative Time Definition and Principal Uncertainity}  \\
\bigskip
\medskip
{\it Martin Sch\"{o}n$^{\dag}$ } \\
\bigskip
{\it Department of Theoretical Physics and Astrophysics}\\
{\it Faculty of Science, Masaryk University } \\
{\it Kotl!!sk! 2, BRNO, Czech Republic }
\end{center}
\newcommand {\beq}{\begin{equation}}
\newcommand {\eeq}{\end{equation}}
\vspace{2cm}
\begin{center}
\end{center}
\vspace{2cm}
\centerline{ABSTRACT}
\begin{quote}
Arguments are given that time must be defined in an operative manner,i.e.,
by constructing devices which can serve as clocks.The investigation of such
devices leads to the conclusion that there is a principal uncertainity of time
if one considers periods which are not large compared with the Planck time.
Thus,according to the old (classical) concept,time cannot be well-defined
at this scale.The uncertainity of time leads to a breakdown of Special and
General relativity in the Planck regime;the same happens with causality.
We present arguments that the classical
concept of time,which treats t simply as a real parameter,must be replaced
by a new one.
\end{quote}
\vspace{1cm}
\dag)e-mail address:schoen\%elanor.sci.muni.cs@csbrmu11.bitnet
\newpage
\ \ \ \
\section{Introduction}
There have been many attempts to quantize gravity but so far there is no
well-defined,satisfying theory of Quantum Gravity.Such a theory,however,is
desirable because the classical general relativistic description breaks down at
the very beginning of our Universe or,e.g.,at the final stage of the formation
of a black hole.It is also required if we agree that the aim of physics is to
provide an uniform picture of nature.\\

Gravitation is closely connected with the structure of space and time,and it
seems that the role of time is crucial for each approach to Quantum
Gravity and should hence be clarified before trying to construct the quantized
theory.There are indications that the present treatment of time is not
adequate.Arguments for that,as far as the canonical quantization approach is
concerned,can be found in Ref.1.Also,the constraints of general relativity
generate dynamical trajectories which in general do not admit a global time
function (see e.g. Ref.2).These observations raise the question about the true
nature of time.The old,so far used time concept,which treats time simply
as a parameter (let us call it the classical one),must be subjected to a
critical investigation.\\

In this paper we inquire the existence of a well-defined time quite generally,
i.e.,also in the presence of non-gravitational fields or in the absense of any
field.For that purpose we investigate the construction of clocks which serve to
define time.We call that procedure an {\it operative definition of time}.
"Operative" means that we are able to construct at least in principle some
device which allows us to measure the considered quantity (here time) with
sufficent exactness (i.e.,the uncertainity of t must be small compared with
t).\\

But why must time be defined operatively?Obviously,there are theories which
contain elements which cannot be defined in such a way.If we take e.g. the
state
vector ("wavefunction") $\psi$ of Quantum Theory,there is no device which
defines directly that state vector.However,there is a fundamental difference
between the concept "state vector" and the classical concept of time: the
underlying theory.\\
Each element of any physical theory has to be interpretated,i.e.,there must
exist a prescription how to "map it into reality" by showing that object(or
group of objects) of the
real world which corresponds to the considered element.Let us consider e.g. the
state vector $\psi$.Quantum Theory provides us with the knowledge of how to
construct out of $\psi$ measurable results.It provides us further with methods
how to generate some particular $\psi$ by constructing special devices.But it
does not claim that $\psi$ is directly realized in nature.All we know is the
correspondence between $\psi$ and derived measurable results.Each of those
(like e.g.the position of a flash) can be defined operatively.Thus we know
how to map $\psi$ into reality,and,hence,there is no problem of definition
present.Let us now consider time.According to the classical concept
(= underlying theory),we can construct out of
the time t essentially just one measurable result,namely t itself.Moreover,in
the case of time the underlying theory (tacitly) claims the t is directly
measurable.Hence,either it must be possible to define t operatively or we have
to change the underlying
theory,i.e.,replace the classical time concept by a new one.\\

In this paper we are going to show that the time cannot be well-defined
operatively if the considered time period is not large compared with the Planck
time.At that
scale a principal uncertainity is showing up which is of the same order as the
time period.This resuld holds independently of the physical situation of the
clock's neighbourhood,i.e.,independent of the presence of fields.We will arrive
at that conclusion by studying various types of possible clocks and
investigating their behaviour taking into account that they have to be material
objects which obey physical laws.We will see that Heisenberg's Uncertainity
Principle and the existence of a Schwarzschild horizon cause limitations
concerning the size of the clock.These limitations lead directly to our
result.\\

{}From these arguments we can infer that we need a new concept of time which
provides us with some framework how to construct out of it results also
measurable at the Planck scale $(\sim 10^{-44}sec)$.The old time concept has
too poor structure to fit to reality at that scale.\\

Additionally,we are going to investigate implications of this result upon
Special and General Relativity (as well as upon causality).We will see that
all break down at Planck scale.That result is not new (at least as far as
General Relativity is concerned);it can be obtained also by considering metric
fluctuations (see e.g. Ref. 3) or,alternatively,by an analysis of measurability
of field quantities similar to that performed by Bohr and Rosenfeld
(cf. Ref. 4).The derivation presented in this paper is from the mathematical
point of view especially simple;all we need are some primitive estimations.\\

However,the main goal of that paper is not to give a new argument that
something fundamental must happen at Planck scale but to investigate the
conceptual nature of time and to motivate trying approaches to Quantum Gravity
with some fundamentally changed time concept.\\
The plan of this paper is as follows.In section 2 we study the possibility of
defining time in an operative manner by considering elementary clocks.
In section 3 we discuss some immediate consequences concerning Special
Relativity and causality.In section 4 we
investigate the principle of equivalence in the light of the results obtained
in the preceeding sections.Section 5 concludes this paper.\\
\\\\
\section{An Operative Definition of Time}
Let us start by considering an arbitrary spacetime which is equipped with a
system of ideal {\it light clocks }.Each of these clocks consists of two
spherical "mirror" masses situated in a distance l between which a photon is
oscillating.Every time when the photon is returning to the sphere from which it
first started a counter is turned one unit further.We denote such a time unit
by t.The time
which is recorded by the counter is called {\it time at the position of the
(first) sphere}.\\

The clock works appropriate only if the following conditions are satisfied.
First of all,the photon must be able to leave the first sphere and to reach
the second one.Thus, l must be greater than the Schwarzschild radius of the
sphere.We obtain:
\begin{equation}\\
l\geq\frac{2Gm}{c^{2}}
\label{eq:1}
\end{equation}\\
where G is the gravitational constant,c is the velocity of light,and m is the
mass of one sphere(let both have equal masses).\\
One can even strenghten that condition by replacing "$\geq$" by "$\gg$";then
we obtain a weak field condition which must be satisfied if we demand that the
gravitational field between the spheres shall be weak.But for our purpose (1)
is sufficient.Note that (1) is only a necessary condition;to be a sufficient
one would require additionally,that the radius of the sphere is greater than
the Schwarzschild radius.Again,the above form is good enough for the
following.\\

Uncertainity of position and momentum of the mirror spheres cause an
uncertainity $\Delta$t in the time t defined by such a clock.In order to have
a "reasonable" definition of time we should demand:
\begin{equation}\\
\Delta t \ll t = 2l/c
\label{eq:2}
\end{equation}\\
(\ref{eq:2}) is equivalent to
\begin{equation}\\
\Delta l \ll l
\label{eq:3}
\end{equation}
We are now going to calculate the minimal uncertainity $\Delta l_{min}$ in l
which is compatible with Heisenberg's Uncertainity Principle.For the
uncertainity $\Delta p$ in the momentum p of the mirror sphere holds:
$\Delta p \approx m\Delta v$,
where v is the velocity of that mass.Since $\Delta v$ causes an uncertainity
$\Delta l$ during the time intervall t which is equal to $t\Delta v$,we get
(according to $(\Delta l\Delta p)_{min} \approx \hbar$)
\begin{equation}\\
\frac{m(\Delta l_{min})^{2}}{t} \approx \hbar
\label{eq:4}
\end{equation}\\
where $\hbar$ is the Planck constant.Since $t = 2l/c$ we obtain finally
\begin{equation}\\
\Delta l_{min} \approx \sqrt{\frac{2\hbar l}{mc}}
\label{eq:5}
\end{equation}
This could be made arbitrarily small by $m \rightarrow \infty$ but we have to
bear in mind condition (\ref{eq:1}).\\

(\ref{eq:3}) together with (\ref{eq:5}) implies
\begin{equation}\\
m \gg \frac{\hbar}{lc}
\label{eq:6}
\end{equation}
(\ref{eq:6}) together with condition (\ref{eq:1}) yields:
\begin{equation}\\
l^{2} \gg \frac{\hbar G}{c^{3}} = l_{Planck}^{2}
\label{eq:7}
\end{equation}
where $l_{Planck}$ is the so-called Planck length (cf.e.g. Ref. 3).From
(\ref{eq:7}) we can infer that the size of our clock must be large compared
with the Planck length.
Since t = 2l/c we obtain finally(neglecting factors like 2):
\begin{equation}\\
t \gg \sqrt{\frac{\hbar G}{c^{5}}} = t_{Planck}
\label{eq:8}
\end{equation}
Thus,the time unit t defined by an ideal clock must be large compared with the
Planck time $t_{Planck}$,if t shall be defined accurately.\\

However,we have considered just one special clock.Could it not be possible
that some appropriately constructed clock enables us to define a time unit
which is smaller than that of (\ref{eq:8})?In order to answer this question
we consider now more general clocks.\\
Let us first investigate an arbitrary clock consisting of some device of size l
and mass m.Let time again be defined by some light ray which is moving
periodingly within that device.Again,we have the condition (\ref{eq:1});
otherwise the clock cannot work.Moreover,the estimations based on Heisenberg's
Uncertainity Principle leading to condition (\ref{eq:6}) are still valid.Thus,
we obtain again our result $t \gg t_{Planck}$.\\
One might also consider some device which is operating with particles of
velocity v instead of photons ("mechanical clock").Then condition (\ref{eq:1})
is still valid(it
can even replaced by a stronger condition,but we do not need that strong
condition).In condition (\ref{eq:6}) c is replaced by v;thus we get instead
of (\ref{eq:7}) $l^{2} \gg \hbar G/c^{2}v$.But since $1/v \geq 1/c$,we get
again the result (\ref{eq:7}),and,hence,also (\ref{eq:8}).\\
So far,we have considered clocks which work according to the same principle
as our ideal light clock.Let us now study clocks which might be really
different from those considered above.We are going to consider a "radioactive"
clock.That is a device consisting of a sample of some material which
decays with half-life time $t_{H}$.That sample is situated in the center of a
hollow sphere which serves to collect all emitted photons.The recorded rate
can be used for a definition of time.If we equip this hollow sphere with
sufficient mass we can decrease its position uncertainity as much as desired.
Hence,it seems that in such a manner it is possible to construct an unlimited
definition of time.But let us consider that clock in greater detail.\\
We assume that the material consists of two-energy-levels systems.The energy
levels are denoted by $E_{1}$ and $E_{0}$,respectively.For such a situation
the well-known relation
\begin{equation}
t_{H}\Delta E \geq \hbar
\end{equation}
means,that if the system has a half-lifetime $t_{H}$,then the difference
$E_{1} - E_{0}$ can be at best determined with an uncertainity $\Delta E$.
An upper boundary for $\Delta$E is the total energy of the sample.
Hence,we obtain:
\begin{equation}
\Delta E \leq mc^{2}
\end{equation}
where m is the sample mass.Let the sample have the size l.Then,again condition
(1) holds,for otherwise the emitted photon would be unable to reach the hollow
sphere.On the other hand,if we denote the minimal time period which can be
measured by our clock by t,we get the condition $l \ll ct$ ,for otherwise the
point where the decay took place is such uncertain that we cannot infer from
the recorded rate the time within the exactness given by t.Namely,if the size
of the sample is of the same order as the distance travelled by the photon
within that time t,the non-centralization of the decay events cause
perturbations
of the recorded rate of order t.(9) and (10) together with that condition
yield:
\begin{equation}
tt_{H} \gg \frac{G\hbar}{c^{5}}
\end{equation}
But since the half-lifetime $t_{H}$ cannot be much bigger than the minimal
measurable time unit t (otherwise the clock would not be exact enough to
detect t),
we obtain again result (8).Thus,also the "radioactive" clock fails to define
time units which are not large compared with the Planck time.\\

In this section we have considered different types of clocks.All of them
showed a principal uncertainity of the time defined thereby if the time unit is
not large compared with the Planck time.Of course,we have not considered all
possible clocks.But since the clocks investigated in this section are very
elementary,it is quite reasonable to assume that the principal uncertainity is
a fundamental
one and a basic feature of nature,and is hence present for all clocks.\\
\\\\
\section{Consequences concerning Special Relativity and Causality}

Special Relativity is based on the behaviour of systems of synchronized clocks,
which form a reference frame.
Let us therefore investigate under which conditions such a synchronization
can be performed sucessfully.\\
Consider two clocks which are separated from each other by a distance
d.Let us try to synchronize them by sending some light signal from one clock
to the other one.Heisenberg's Uncertainity Principle leads again to some
minimal uncertainity $(\Delta d)_{min}$ in the distance which is approximately
equal to
$\sqrt{\frac{\hbar d}{mc}}$ (see (\ref{eq:5}).In order to be able to perform
such a synchronization procedure the distance d must exceed the Scharzschild
radius
of the clock,i.e.,$Gm/c^{2} \leq d$.Putting both results together we obtain
$(\Delta d)_{min} \geq \sqrt{\hbar G/c^{3}}$ (neglecting factors like 2).
Thus,the minimal uncertainity in
d is at least equal to the Planck length.This causes an uncertainity in the
synchronization $(\Delta t)_{syn}$ which is equal to $(\Delta d)_{min}/c$.
Thus,we get the result
\begin{equation}\\
(\Delta t)_{syn} \geq t_{Planck}
\label{eq:10}
\end{equation}
Again,the Planck time turns out to be the fundamental limit:it is not possible
to synchronize clocks more perfectly than $t_{Planck}$.\\
We can conclude that Special Relativity is limited by the
condition that the considered times are large compared with the Planck time.The
reason is the principal synchronization uncertainity (\ref{eq:10}):the
possibility
to synchronize a clock frame breaks down when we approach the Planck time.The
concept {\it time of a reference frame} becomes senseless.\\

Now let us study what happens with causality at Planck scale.Two states are
said to be causally
connected if the first one,say A,taking place at time $t_{A}$ determines the
second one called B,taking place at time $t_{B}$.The concept of causality is
quite unproblematic in classical physics,but can also be maintained in
Relativity and Quantum Theory,if the concept of state is defined appropriately.
However,as far as the latter is concerned,this is not completely true.There is
one process which violates causality:the reduction of the wavefunction due to
some measurement.But even in that case it is possible to save the causal
relation
if we give up determinism and replace it by some weak causal principle which
states that A causes B with some probability.\\
In any case,if A and B are related by a causal relation,then $t_{A}$ cannot be
bigger than $t_{B}$.Now,let us approach the Planck scale.If $t_{B} -t_{A} \sim
t_{Planck}$,it is impossible to say which event happens first.It becomes
senseless
to say that one causes the other.On the other hand,it is also not reasonable to
say that they cause each other mutually,for there is still some asymmetry
between A and B:unless $t_{B} - t_{A} = 0$, $t_{B}$ is more often be observed
to occur "later" (according to necessarily unsharp clocks) as $t_{A}$.
However,with some new time concept that asymmetry could appear in a new light;
and perhaps it is then possible to reformulate the causal principle in such a
way that it holds also for the Planck scale.But according to the present
concept we must conclude that causality breaks down or looses its sense in
the Planck regime.\\
\\\\
\section{\bf Consequences concerning Gravity}

In this section we will investigate whether the equivalence principle which
is essential for General Relativity can be applied if the gravitational field
is such strong that it changes significantly across a distance which is not
large
compared with the Planck length.Such a situation was realized near Big Bang
when the age of the Universe was of order of magnitude of the Planck time.\\
The equivalence principle tells us that it is possible to find space-time
coordinates such that locally Special Relativity holds.One can formulate this
principle in greater detail and distinguish between strong and weak principle
but for our purpose the above form is sufficient.Let us now try to apply this
principle to some region of a strong gravitational field.The region which can
be described locally in terms of Special Relativity by choosing appropriate
coordinates must be smaller than the size of significant change of the field
strength.Since we are considering the case that this size is of order of
magnitude of the Planck length our local neighbourhood cannot be equipped with
clocks of size much larger than that length.In other words,Einstein's comoving
box is to small for accurate clocks which are necessarily much larger than
$l_{Planck}$ ,as we have seen in section 2 (see eq. (7)).\\
Let us now imagine some hypothetical observer within that comoving box.He can
use only very small clocks which are necessarily very uncertain.Moreover,
according to our investigations performed in the previous section,for such
small time units Special Relativity breaks down.It is thus meaningless to
say that for our observer the laws of Special Relativity hold.Hence,the
equivalence principle cannot be applied near the Planck regime.\\

But without beeing able to apply the equivalence principle General Relativity
cannot be an appropriate description of strong gravitational fields.That is
a well-known result which is also confirmed by arguments different from those
considered in this paper(see Ref.3,4 ).Since,as we have seen above,this
breakdown
is connected with the rather serious impossibilty of maintaining the
"classical"
concept {\it time at one point} we should first replace that concept by a new
one before trying to quantize gravity.That new,so far unknown concept must be
one which reflects the principal uncertainity of time in the Planck regime.\\

Let us end this section with a very undetailled proposal about the construction
of a future
theory which is able to decribe gravity in the Planck regime and,hence,the
very early universe.The proposal is quite a conservative one for it tries to
preserve the essence of the equivalence principle.It consists in
performing the following steps:\\
\begin{enumerate}
\item Find some generalization of Special Relativity which is valid also for
short
time periods.For that purpose a new time concept is required which reveals
a principal and unescapable uncertainity near Planck time.Note that it is not
sufficient to replace t by some operator $\hat{t}$ and to construct in such a
way some "quantum theory of time" because then we can choose an eigenstate of
that operator and get an exact time value.But we saw that time alone and for
itself becomes uncertain.\\
\item Apply a slightly modified equivalence principle which states that it is
possible to find coordinates such that locally the theory to be constructed
in the first step holds.\\
\item Try to quantize that new general relativistic theory.\\
\end{enumerate}

Perhaps,there are close relations between step 1 and step 3 such that they
must be performed together before applying the equivalence principle.But
these are speculations which we do not want to continue now.\\
\\\\
\section{\bf Conclusions and Outlook}

We have seen that an operative definition of time is not possible when we
are considering time periods which are not large compared with the Planck
time ($\sim 10^{-44} sec$).Any possible clock will become uncertain and unsharp
when we decrease its size down to the Planck length such that the thereby
defined time unit becomes comparable with the Planck time.Thus,we must give
up the concept of a continuous,exactly measurable time and have to replace
it by a new time concept which reflects this principal uncertainity for
small periods.Note that this new concept cannot simply be some replacement
of the classical parameter t by an operator $\hat{t}$ for in this case the
measured value could be made arbitrarily exact by choosing some eigenstate
of $\hat{t}$.The principal time uncertainity is more serious and requires
a mathematical framework which is different from the ordinary operator
language of the present quantum theory.\\
Note also that this principial uncertainity is not merely a problem of time
measurement.We have adopted the point of view that if a physical quantity
is present in the underlying theory then it must be possible in principle
either to measure this quantity directly or to construct out of it
measurable results.The latter is e.g. true in Quantum Theory,where the
state vector cannot be "verified" directly in nature by some suitable
device.However,Quantum Theory tells us how we can derive from the state
vector operatively defined quantities.On the other hand,the classical "time
theory" tells us (at least tacitly)
that time,contrary to the state vector,can be measured directly.According to
this concept,time is an "elementary" quantity.Hence,it must be possible to
define it operatively.Thus,the observed principal
uncertainity at Planck scale shows that the classical time is not
well-defined.\\

This principal time uncertainity shows up regardless whether some field is
present or not.Thus,the Planck regime requires fundamental changes not only
for strong gravitational fields but also for all other fields as well as
for flat spacetimes without any field.Moreover,we have seen that Special
Relativity breaks down near the Planck regime also due to a second reason:
it is then impossible to synchronize a system of clocks forming a frame of
reference and hence to define the common time of that system.We have further
considered the consequences of that principal uncertainity concerning
causality.
We found,that there is no well-defined causal relation at Planck scale.The same
happens with General
Relativity.We saw that for strong gravitational fields which change
significantly
across
some distance which is not large compared with the Planck length the
equivalence principle cannot be applied.Hence,General Relativity must break
down at Planck regime;a result which is also confirmed by other observations
(cf. Refs. 3,4).\\
However,there is some hope that a generalisation of these theories can be
found which is also valid in the Planck regime if we construct an appropriate
new time concept.Perhaps,such a concept could be the clue to find
some satisfying theory of Quantum Gravity.We believe that the time problem
is the key problem for constructing such a theory.\\

Of course,all our results are based on the assumption that Heisenberg's
Uncertainity Principle as well as condition (1) arising from the existence
of a Schwarzschild horizon are at least approximately valid (i.e.,sufficient
for our estimations performed in section 2).
If this is not true,then perhaps the
classical concept of time could be maintained but at least one of these
basic laws must be altered fundamentally in the Planck regime.However,it
seems to be quite reasonable to assume that the estimations performed in
the previous sections are justified.In this case,the classical concept of
time cannot be maintained.\\
\\\\
{\bf Acknowledgement}\\

The author wishes to express his thanks to the hospitality of the Department
of Theoretical Physics and Astrophysics,Masaryk University Brno,Czech Republic,
where the main part of this work has been carried out.\\
\\\\
\bigskip
\\\\
\medskip
{\bf References}\\
\begin{enumerate}
\item  K.V.Kuchar in {\it Quantum Gravity 2},edited by C.Isham,R.Penrose,and
D.W.Sciama (Clarendon,Oxford,1981)
\item  P.Hajicek,Phys. Rev. D {\bf 34},1040 (1986)
\item  J.A.Wheeler,Ann. Phys. (N.Y.){\bf 2},604 (1957)
\item  B.S. de Witt in {\it Gravitation: An Introduction to Current Research},
ed. by L.Witten (John Wiley \& Sons,Inc.,New York,1962)
\end{enumerate}
\end{document}